\documentclass[12pt]{article}
\usepackage{amsfonts,amssymb,amsmath,amscd}

\newcommand{\ZZ}{{\mathbb Z}}
\newcommand{\RR}{{\mathbb R}}
\newcommand{\CC}{{\mathbb C}}
\newcommand{\op}{\oplus}

\newcommand{\bff}{{\bar f}}

\newcommand{\om}{{\omega}}

\newcommand{\fone}{{F_{1,1}}}
\newcommand{\fzero}{{F_{0,2}}}
\newcommand{\ftwo}{{F_{2,0}}}
\newcommand{\gi}{g^{-1}}
\newcommand{\gt}{g^t}
\newcommand{\gti}{{g^t}^{-1}}
\newcommand{\fonet}{F^{\,t}_{1,1}}

\title{Stability Conditions For Topological D-branes: A Worldsheet Approach}
\author{Anton Kapustin, Yi Li \\{\it California Instutute of Technology, Pasadena, CA 91125, U.S.A.}}

\begin{document}

\begin{titlepage}

\maketitle

\begin{abstract}
We study conditions on the topological D-branes of types A and B obtained by requiring a proper matching of
the spectral flow operators on the boundary. These conditions ensure space-time supersymmetry
and stability of D-branes. In most cases, we reproduce the results of Marino-Minasian-Moore-Strominger,
who studied the same problem using the supersymmetric Born-Infeld action. In some other cases, 
corresponding to coisotropic A-branes, our stability condition is new. 
Our results enable us to define an analogue of the Maslov class and grading for coisotropic A-branes. 
We expect that they play a role in a conjectural generalization of the Floer homology. 

\end{abstract}

\end{titlepage}

\section{Introduction}

Supersymmetric D-branes play an important role in string theory and its mathematical applications, such
as mirror symmetry. There are two rather different approaches to deriving the conditions for space-time supersymmetry
in the presence of D-branes. The first approach uses the space-time viewpoint, where D-branes are regarded
as ``solitons,'' whose low-energy dynamics is described by gauge fields, scalars, and fermions. 
One can deduce the conditions of unbroken space-time supersymmetry by studying the supersymmetric Born-Infeld
action for these fields.
The second approach uses the worldsheet viewpoint, where D-branes are regarded as boundary conditions for open
strings. The condition of space-time supersymmetry is equivalent to the requirement that the boundary condition
preserve a suitable chiral algebra. 

The case which has been studied most thoroughly is when the gauge field on the D-brane is flat. In this case
it is well-known that the two methods give equivalent results~\cite{BBS,OOY,Betal}. In the case when the gauge field on the D-brane
is not flat, the conditions of space-time supersymmetry have been analyzed in Ref.~\cite{MMMS} using the
Born-Infeld action. 

In this paper we derive the supersymmetry conditions for D-branes from the worldsheet point of view. We limit
ourselves to the case when the relevant worldsheet chiral algebra is an $N=2$ superconformal algebra plus a spectral flow
operator. This corresponds to half-BPS D-branes on manifolds of $SU(n)$ holonomy. 

There are several reasons why we
think it is worthwhile to revisit this problem. First, the worldsheet derivation allows one to obtain
the conditions of supersymmetry in all dimensions at once, while in the Born-Infeld approach one has to deal
with spinors, whose properties depend on the dimension in a rather complex way. Second, in the analysis of Ref.~\cite{MMMS}
certain interesting cases have been missed. For example, it is possible to have a supersymmetric 5-cycle on
a complex 3-fold if the gauge field on the cycle is not flat. 

More generally, given a Calabi-Yau $X$, there may
exist supersymmetric D-branes of type A which are not special Lagrangian submanifolds and carry non-flat gauge fields.
The conditions on such branes coming from the requirement of $N=2$ worldsheet supersymmetry have been worked out
in Ref.~\cite{KO}. In particular, such A-branes must be coisotropic, rather than Lagrangian, submanifolds. However, 
the conditions which ensure the preservation of the spectral flow operator have not yet
been determined for such branes. From the point of view of topological string theory, a boundary condition which
preserves $N=2$ worldsheet supersymmetry is a topological D-brane, while the preservation of the spectral flow
operator is a {\it stability} requirement. Thus a supersymmetric D-brane is the same as a stable topological D-brane.
But stability is important even if one stays within the confines of topological string theory. To explain this, recall
that on the classical level every Lagrangian submanifold $Y$ of a Calabi-Yau $X$ gives rise to a topological A-brane.
On the quantum level, this is not true because of anomalies in the worldsheet R-charge. In order for anomalies
to vanish, a certain class in $H^1(Y,\ZZ)$, the Maslov class, must vanish~\cite{Hori}. The definition of the Maslov class
is closely related to the notion of a special Lagrangian submanifold; in particular, it is easy to show that any special
Lagrangian submanifold has zero Maslov class. On the other hand, the condition of being ``special'' is precisely
the stability condition for Lagrangian A-branes. For coisotropic A-branes, the conditions of anomaly cancellation
are not known, but we expect that they are related to stability conditions, in the same way as the vanishing of Maslov
class is related to the condition of being ``special.'' Using this intuition, we propose in this paper an analogue
of the Maslov class for coisotropic A-branes of Ref.~\cite{KO}. We also propose a coisotropic analogue of the notion
of a ``graded Lagrangian submanifold'' introduced in Ref.~\cite{Konts}.

The outline of the paper is as follows. In Section~\ref{Bbranes} we derive the stability conditions for D-branes
of type B. Our results are in complete agreement with Ref.~\cite{MMMS}. In Section~\ref{Abranes} we derive the 
stability conditions for D-branes of type A, allowing the gauge field on the branes to be non-flat. In Section~\ref{disc}
we discuss our results and propose definitions of the generalized Maslov class and grading for 
coisotropic A-branes. 

\section{Stability conditions for B-branes}\label{Bbranes}

Let $X$ be a K\"ahler manifold of dimension $n$ with a trivial canonical line bundle. We will denote by $\Omega$ a non-vanishing
holomorphic $n$-form on $X$; it is unique up to a constant multiple. The metric tensor on $X$ will be denoted $G$.
Let $Y$ be a submanifold of $X$ carrying a line bundle $E$ with a unitary connection $\nabla$. Let $F$ be
the curvature 2-form of $\nabla$. It is a real-valued 2-form on $Y$. The boundary condition for the
fermions has the form
$$
\psi_+=R\psi_-,
$$
where $R$ is an orthogonal transformation of $TX\vert_Y=NY\op TY$ given by
\begin{equation}\label{R}
R=\begin{pmatrix} -1 & 0 \\ 0 & (g-F)^{-1}(g+F) \end{pmatrix}.
\end{equation}
Here $g$ is the restriction of the K\"ahler metric to $Y$.
The boundary condition for bosons can be determined by requiring $N=1$ supersymmetry, but we will not need it
here. By definition, a D-brane of type B must preserve the sum of the left and right $N=2$ super-Virasoro generators.
This requirement imposes a constraint on $Y$ and $(E,\nabla)$: $Y$ must be a complex submanifold of $X$, and
$F$ must have type $(1,1)$. Thus $(E,\nabla)$ is a holomorphic line bundle. We will denote by $k$ the complex
dimension of $Y$. Note that as a consequence of these
requirements $R$ commutes with the complex structure tensor $I$. This can be seen more directly by recalling that
the R-currents have the form
$$
J_\pm\sim G(\psi_\pm,I\psi_\pm).
$$
Since $R$ is orthogonal and takes $J_+$ to $J_-$, it must commute with $I$.

The spectral flow operators have the form
$$
S_\pm=\Omega(\psi_\pm,\psi_\pm,\ldots)=\frac{1}{n!}\Omega_{i_1\ldots i_n}\psi_\pm^{i_1}\ldots\psi_\pm^{i_n}.
$$
The matching of the spectral flow operators on the boundary requires
$$
S_+ =e^{i\alpha} S_-,
$$
where $\alpha\in \RR$ is a constant. Let $R_h$ be the holomorphic part of $R$, i.e. the part of $R$ which maps
$TX^{1,0}$ to $TX^{1,0}$. Then the above condition becomes
$$
\det R_h=e^{i\alpha}.
$$
Taking into account Eq.~(\ref{R}) and introducing the K\"ahler form $\omega=GI$, we can rewrite this
condition in the following form:
$$
\det(\omega\vert_Y+iF)=e^{i\alpha+i (n-k)\pi}\det(g+F).
$$
where $g$ is the restriction of the metric to $Y$.
In terms of differential forms, we can write this as follows:
$$
\frac{1}{k!}(\omega\vert_Y+iF)^{\wedge k}=i^{n-k} e^{i\alpha/2}{\sqrt{\det(1+g^{-1}F)}}\ {\rm vol}_Y.
$$
This condition agrees with the stability condition
derived in Ref.~\cite{MMMS} from the Born-Infeld action. For small $F$ it reduces to the Donaldson-Uhlenbeck-Yau
equation:
$$
F\wedge\omega^{k-1}=c\cdot \omega^k,\quad c\in \RR.
$$

\section{Stability conditions for A-branes}\label{Abranes}

By definition,  a D-brane of type A is a boundary condition which preserves the sum of the left-moving $N=2$
super-Virasoro and the mirror of the right-moving $N=2$ super-Virasoro. Since the mirror automorphism maps the R-current
to minus itself, this implies that the reflection operator $R$ must map $J_+$ to $-J_-$. Since $R$ is orthogonal,
this means that it anti-commutes with $I$, i.e. it maps $TX^{1,0}$ to $TX^{0,1}$, and vice versa. 

On the classical level, the necessary and sufficient conditions for a D-brane to be an A-branes have been determined in 
Ref.~\cite{KO}. These conditions are slightly more complicated than for B-branes. The first requirement is that $Y$ 
must be a coisotropic submanifold of $X$. This means that ${\rm ker}\,\om|_Y\equiv TY^\om\subset TY $ forms an integrable 
distribution 
of constant rank in $TY$. Denote the quotient bundle $TY/TY^\om$ by $FY$. The second requirement then says that the curvature 
2-form 
$F$ of the line bundle annihilates $TY^\om$ and therefore descends to a section of $\wedge^2 FY^*$. Finally, for the 
boundary condition to be a topological A-brane, the endomorphism $\om^{-1}F|_{FY}$ should be a complex structure on $FY$. 
It follows from these conditions that the complex dimension of $FY$ is even~\cite{KO}. We denote the complex dimension of $FY$ to
 be $2k$.
If $Y$ is of real codimension $r$ in $X$, then the complex dimension of $X$ is $n=r+2k$.

We use the metric to decompose the bundle $TX\vert_Y$ as a direct sum:
$$
TX\vert_Y\simeq NY\op TY^\om \op FY.
$$
We can choose a local complex basis $e_1,\ldots,e_n$ for $TX$ such that $e_1,\ldots,e_r,$ and their complex-conjugates
span $NY\op TY^\om$, and $e_{r+1},\ldots,e_n$ and their complex-conjugates span $FY$. The dual basis for $TX^*$ will be denoted
$f^1,\ldots,f^n$.
In such a basis the ``holomorphic'' part of $R$, i.e. 
the part which maps $TX^{0,1}$ to $TX^{1,0}$, can be represented by a matrix
\begin{equation}\label{RA}
R_{\rm h}\;=\;\begin{pmatrix} 1_{r\times r} & 0 \\ 0& (G-F)^{-1}(G+F)\big|_{FY^{0,1}} \end{pmatrix}
\end{equation}
The lower right block can be simplified by writing
$$G|_{FY} = \begin{pmatrix}0&g\\ \gt&0\end{pmatrix},\qquad F|_{TY} = \begin{pmatrix}\ftwo&\fone\\ -\fonet&\fzero\end{pmatrix}.$$
Then the condition $(\om^{-1}F|_{FY})^2 = -1$ shows that both $\ftwo$ and $\fzero$ are nondegenerate, and they satisfy
\begin{equation}
	\begin{array}{l}
	\gi\ftwo \;=\; -(1+\gi\fone)(\gti\fzero)^{-1}(1+\gti\fonet)\\
	\gti\fzero \;=\; -(1-\gti\fonet)(\gi\ftwo)^{-1}(1-\gi\fone)\end{array}
\label{eq:F_cond}
\end{equation}
From this it follows that
\begin{equation}\label{GF}
(G-F)^{-1}(G+F)\big|_{FY^{0,1}} \;=\; \gti\fzero(1-\gi\fone)^{-1}.
\end{equation}

The matching conditon for the spectral flow operator on the boundary is
$$S_+ =e^{i\alpha}\, \bar{S}_-,$$
where again $\alpha\in\RR$ is a constant. Writing out in components and making use of (\ref{RA}),(\ref{GF}), this becomes
\begin{equation}\label{Ommatch}
\Omega_{12\ldots n}\,{\rm det}\,\fzero = e^{i\alpha}\,\bar{\Omega}_{12\ldots n}\,{\rm det}\,(g-\fone).
\end{equation}
Recall that the holomorphic top form $\Omega$ satisfies $|\Omega_{12\ldots n}|^2 =c\cdot \sqrt{\det G}$ where $c\in \RR$ is
a constant. Multiplying Eq.~(\ref{Ommatch}) by 
$\Omega_{12\ldots n}$, we get
$$
(\Omega_{12\ldots n})^2\,\det\fzero \;\sim\; e^{i\alpha}\sqrt{\det G}\cdot\det\,(g-\fone).
$$
The right hand side can actually be expressed in terms of the full metric tensor, by noting that
$$\frac{\det\,(G+F)}{\det\,G} \;=\; 2^{2k}\,\det(1-\gi\fone).$$
To derive this identity, one has to use Eq.~(\ref{eq:F_cond}).
Thus the spectral flow matching condition can be written as
\begin{equation}
(\Omega_{12\ldots n})^2\,{\rm det}\,\fzero \;\sim\; e^{i\alpha}\cdot\frac{\det\,(G+F)}{\det\, G} {\sqrt{\det\, G\, \det\, G|_{FY}}}
\label{eq:spec_A}
\end{equation}

To see the geometric meaning of Eq. (\ref{eq:spec_A}), we note that one may always assume that $NY$ is spanned by the imaginary
parts of $e_i,i=1,\ldots,r,$ and $TY^\om$ is spanned by the real parts of $e_1,\ldots,e_r$. 
Consider the following differential form on $Y$:
\begin{multline}\label{restriction}
\frac1{k!}\Omega\vert_Y\wedge F^{\wedge k} =\\
 \Omega_{12\ldots n}\,{\rm Pf}(\fzero)\cdot {\Re f^1}\wedge\cdots\wedge {\Re f^r}\wedge f^{r+1}\wedge \ldots\wedge f^n\wedge 
\bff^{r+1}\wedge\cdots\wedge \bff^n, 
\end{multline}
where ${\rm Pf}(\fzero)$ is the Pfaffian of $\fzero$. We can rewrite this as follows:
\begin{equation}
\frac1{k!}\Omega\vert_Y\wedge F^{\wedge k} \sim 
\frac{\Omega_{12\ldots n}\,{\rm Pf}(\fzero)}{\sqrt{{\rm det}\,G|_Y}}
\cdot {\rm vol}(Y)
\label{eq:vol}
\end{equation}
Comparing (\ref{eq:spec_A}) with (\ref{eq:vol}), and noting that 
$$
\det\fzero={\rm Pf}(\fzero)^2,\quad \det\, G \det\, G|_{FY}=\left(\det\, G\vert_Y\right)^2,
$$
we obtain
\begin{equation*}
\frac1{k!}\Omega\vert_Y\wedge F^{\wedge k} \sim e^{i\alpha/2}\sqrt{\frac{\det\,(G+F)}{\det G}}\cdot {\rm vol}(Y)
\end{equation*}

Let us compare this condition with known results. If the bundle on the brane is flat, $F=0$, then $Y$ is an A-brane if and only
if it is Lagrangian. The above stability condition then becomes
$$
\Omega\vert_Y\sim e^{i\alpha/2}{\rm vol}(Y).
$$
This means that $Y$ is a special Lagrangian submanifold, in agreement with Refs.~\cite{BBS,OOY}. For $F\neq 0$, we can compare 
with the results of Ref.~\cite{MMMS}.
The physical range for $n$ is $n=2,3,4$. For $n=2$ $k=1$ (a 4-brane wrapped on a 2-fold) Marino et al. found a family of 
supersymmetric D-branes parametrized by a $U(2)$ matrix. A-branes correspond to the case when this matrix has zero
elements on the diagonal. Then their stability condition reads
$$
\Omega\wedge F\sim e^{i\beta} \sqrt{\frac{\det\,(G+F)}{\det G}}\cdot {\rm vol}(Y), \quad \beta\in \RR
$$
in complete agreement with our result. For $n=4$ we can have $k=1$ or $k=2$, i.e. a 6-brane or an 8-brane. (A Lagrangian 4-brane in
a complex 4-fold is also possible, but then $F=0$). One can check that the conditions on the 6-brane found by Marino et al.
agree with ours. We could not compare the conditions on the 8-brane, because in Ref.~\cite{MMMS} they are stated only for B-branes.

The remaining case $n=3$, $k=1$ (5-brane in a complex 3-fold) has not been analyzed before. Stable A-branes of this kind are possible 
only if
$b_5(X)\neq 0$, which means that $X$ is either a complex torus of dimension $3$ or a product of a $T^2$ and a K3-surface.

\section{Discussion}\label{disc}

The main result of this paper is the determination of the stability condition for A-branes in the case when $F\neq 0$. 
This condition can be reformulated in different ways. One convenient formulation, equivalent to the one given above, is that
$$
{\rm Im}\left(e^{-i\alpha/2}\Omega\vert_Y\wedge F^k\right)=0
$$
Here $k$ is half the complex dimension of the bundle $FY$. We would like to emphasize that our analysis (as well as most other
existing derivations of supersymmetry conditions for D-branes) applies only when the vector bundle on the brane has rank one.

Analogy with the case of Lagrangian A-branes~\cite{Hori} suggests that our stability condition ensures the absence of anomaly in 
the R-charge conservation for worldsheets ending on the D-brane $(Y,E,\nabla)$. It would be interesting to show this explicitly. 
In any case, our stability condition
suggests a definition of the ``generalized Maslov class'' for coisotropic A-branes. For any A-brane, stable or not, we can
consider the form $\Omega\vert_Y\wedge F^k$. This is a top form on $Y$, and therefore it is equal to the volume form of $Y$ times a
complex-valued function $f$. From Eq.~(\ref{restriction}) and non-degeneracy of $F_{0,2}$ it follows that this function is 
non-zero everywhere 
on $Y$. Then the 1-form $d\log f$ defines
a class in $H^1(Y,\CC)$ which we take as the generalized Maslov class. As usual, one can do slightly better and define a
class in $H^1(Y,\ZZ)$ whose reduction to $\CC$ is the class of $d\log f$. One chooses an open cover of $Y$ such that all double
overlaps are connected, chooses a branch of $\log f$ on each set of the cover, and considers the Cech 1-cocyle with
values in $\ZZ$ which measures the obstruction to the existence of the global $\log f$. Its cohomology class is the
integral version of the generalized Maslov class. 

If the generalized Maslov class vanishes in $H^1(Y,\ZZ)$, one can define the notion of a graded coisotropic A-brane which
generalizes the notion of a graded Lagrangian submanifold defined by Kontsevich~\cite{Konts}. Namely, a graded coisotropic
A-brane is an A-brane together with a global choice of the branch of $\log f$. Presumably, there exists a coisotropic version
of the $\ZZ$-graded Floer complex for any pair of graded coisotropic A-branes. The cohomology of this complex should
compute the space of open topological strings stretched between the two branes. A formal definition of the generalized
Floer homology in terms of a certain sheaf on the space of paths from one brane to the other has been proposed
in Ref.~\cite{KOlect}.

\section*{Acknowledgments}

A. K. would like to thank Nigel Hitchin and Dmitri Orlov for discussions. This work was supported in part by the DOE grant
DE-FG03-92-ER40701.

\end{document}